\documentclass[preprint,showpacs,amsmath,amssymb]{revtex4}
\usepackage{amsmath} %
\usepackage{graphicx}
\usepackage{amsfonts}
\usepackage{dcolumn}
\usepackage{bm}

\begin{document}

\title{An efficient strategy to suppress epidemic explosion in heterogeneous metapopulation networks}

\author{Chuansheng Shen$^{1,2}$}

\author{Hanshuang Chen$^{3}$}

\author{Zhonghuai Hou$^{1}$}\email{hzhlj@ustc.edu.cn}

\affiliation{$^{1}$Hefei National Laboratory for Physical Sciences
at Microscales \& Department of Chemical Physics, University of
 Science and Technology of China, Hefei, 230026, China \\
 $^{2}$Department of Physics, Anqing Teachers College, Anqing, 246011, China \\
 $^3$School of Physics and Material Science, Anhui University, Hefei, 230039, China}


\begin{abstract}
 We propose an efficient strategy to suppress epidemic
explosion in heterogeneous metapopulation networks, wherein each
node represents a subpopulation with any number of individuals and
is assigned a curing rate that is proportional to $k^{\alpha}$ with
$k$ the node degree and  $\alpha$ an adjustable parameter. We have
performed stochastic simulations of the dynamical reaction-diffusion
processes associated with the susceptible-infected-susceptible model
in scale-free networks. We found that the epidemic threshold reaches
a maximum when the exponent $\alpha$ is tuned to be
$\alpha_{opt}\simeq 1.3$. This nontrivial phenomenon is robust to
the change of the network size and the average degree. In addition,
we have carried out a mean field analysis to further validate our
scheme, which also demonstrates that epidemic explosion follows
different routes for $\alpha$ larger or less than $\alpha_{opt}$.
Our work suggests that in order to efficiently suppress epidemic
spreading on heterogeneous complex networks, subpopulations with
higher degrees should be allocated more resources than just being
linearly dependent on the degree $k$.
\end{abstract}


\maketitle

\section{Introduction}
In the last two decades, we have witnessed dramatic advances in
complex networks research, which has been one of the most active
topics in statistical physics and closely related disciplines
\cite{RMP02000047,AIP02001079,SIR03000167}. The central issue in
this field is to study how the network topology influences the
dynamics \cite{PRP06000175,PRP08000093,RMP08001275}. As one of the
typical dynamical processes built on complex networks, epidemic
spreading has attracted more and more significant attention
\cite{PRL01003200,PRE01066117,
PhysRevE.66.047104,PRL03028701,PRL04178701,PRE01066112,SCI01001316,EPJB02000521,
PRE03035103,PRE04066105,PRE04030902,EPL05000315,PRE07016108,PRL10218701,NAP2010888,NJP10093009,PLoS111002174,PhysRevX.1.011001}.
Despite much effort, many aspects of its role in the case of
metapopulation models are still unclear and deserve more
investigation.

Very recently, metapopulation dynamics on heteregeneous networks,
which incorporate mobility over the nodes, local interaction at the
nodes, and a complex network structure, has gained great research
attention
\cite{NAP2007276,PRL07148701,JTB2008450,JTB2008509,PRE08016111,SCI20091071,PhysRevX.1.011001,SCI20101616,
PNAS098429,
ARPC2009449,Grz2009,PRE10046116,JTB201287,PR2011001,EPL201168009,Moi2009,
NAP2011581,NAP201232,NAP2010544}. In this context,
reaction-diffusion (RD) processes have been widely used to model
phenomena as diverse as epidemic and computer viruses spreading
\cite{NAP2007276,PRL07148701,JTB2008450,JTB2008509,PRE08016111,SCI20091071,PhysRevX.1.011001},
biological pattern formation \cite{SCI20101616,PNAS098429}, chemical
reactions \cite{ARPC2009449,Grz2009,PRE10046116}, population
evolution \cite{JTB201287}, and many other spatially distributed
systems \cite{PR2011001,EPL201168009,Moi2009,NAP2011581}. In a
series of important papers, Colizza \emph{et al}. \cite{NAP2007276}
provided an analysis of the basic RD process of the
susceptible-infected-susceptible ($SIS$) model defined on
heterogeneous metapopulation networks. Therein, each network node
represents  an urban area together with its population and edges
represent air travel fluxes along which individuals diffuse,
coupling the epidemic spreading in different urban areas. They paid
particular attention to the epidemic threshold $\rho_c$, and found
that $\rho_c$ is strongly affected by the topological fluctuations
of the network for diffusing susceptible individuals. Later, Balcan
and Vespignani \cite{NAP2011581} extended such analysis to
non-Markovian diffusive processes on complex networks, wherein
individuals have a memory of their location of origin and displaced
individuals return to their original subpopulation with a certain
rate. Very recently, Vespignani \cite{NAP201232} reviewed and
highlighted some of the recent progress in modelling dynamical
processes that integrates the complex features and heterogeneities
of real-world systems. Nevertheless, all the studies so far have
treated the curing rate $\mu$ as a homogenous parameter, i.e., it is
not dependent on the local property of the network node, such as the
degree $k$. Note, however, in reality the curing rate of individuals
should certainly be associated with the available medical resources
in the local subpopulation, i.e., it is reasonable to assume that
$\mu$ is a function of the degree $k$. It is therefore interesting
to ask: how would the metapopulation dynamics of the $SIS$ model,
for instance, the epidemic threshold $\rho_c$, depend on such a
$k$-dependent curing strategy? The answer to this question may
provide useful instructions regarding the control of epidemic
explosion in metapopulation networks.

  In the present paper, we have addressed such a question by
considering a simple strategy, $\mu_k \sim k^\alpha$, where $k$
denotes the node degree and $\alpha$ is an adjustable parameter. If
$\alpha=0$, one recovers the usual cases studied in previous works.
Herein, we mainly focus on the influence of varying $\alpha$ on
$\rho_c$. Interestingly, we found that $\rho_c$ bypasses a clear-cut
maximum at a certain $\alpha_{opt}$, which corresponds to an optimal
strategy to suppress epidemic explosion. This observation along with
the value of $\alpha_{opt}$ is robust to the change of the network
size and the average network degree. To place the finding on a solid
foundation, we have also performed a mean field (MF) analysis,
wherein $\rho_c$ is identified as the onset point where the global
healthy state with no infected individuals loses stability. The MF
equations successfully reproduce the $\rho_c \sim \alpha$
dependences, and also provide more insights regarding the routes to
epidemic explosion for different values of $\alpha$.

\section{ Model Description } \label{sec2}

We consider a system of $N$ distinct subpopulations, each
corresponding to a network node.  Individuals inside each node  run
stochastically through the paradigmatic $SIS$ model
\cite{Dal1999,SIAM2000599,Mur2002}. Schematically, the stochastic
infection dynamics is given by:
\begin{equation}\label{eqModel}
S + I\xrightarrow{\beta }2I,{\kern 10pt} I\xrightarrow{\mu_k}S
\end{equation}
The first reaction reflects the fact that each susceptible
individual becomes infected upon encountering one or more infected
individuals at a probability rate $\beta$. The second indicates that
infected individuals are cured and become again susceptible at a
$k$-dependent rate $\mu_k$. Inside each network node, reaction
processes take place under the assumption of a homogenous mixing and
conserving the total number of individuals. After the reaction,
individuals randomly diffuse along the edges departing from its
local node.

 In this model, a significant and general result is that the system
undergoes an absorbing-state phase transition with density $\rho$
increasing, in analogy with critical phenomena \cite{RMP08001275}.
Here $\rho$ is defined as the total number of individuals divided by
the number $N$ of network nodes. The critical density $\rho_c$
indicates the epidemic threshold, what we are interested in.

To begin, we perform our strategy on scale-free (SF) networks by using the
Barab\'{a}si--Albert (BA) model \cite{SCI99000509} with power-law
degree distribution $p(k)\sim k^{-3}$. Scale-free networks are much
more heterogeneous and serve as better candidates to test our
strategy than other homogeneous networks, such as small-world or
random networks. For a node $i$ with degree $k_i$, the curing rate
is given by
\begin{equation}\label{strategy}
\mu_ {k_i} = \frac{{k_i ^\alpha  }} {{\sum\nolimits_j {k_j ^\alpha/N
} }}
\end{equation}
Herein, $\mu_{k_i}$ is normalized such that the average curing
rate remains constant: $\bar \mu = \frac{1}{N}\sum_i \mu_{k_i}=1$.
Note that in other related works about epidemic dynamics on networks , a $k$-dependent strategy, but associated with the infection rate, had also been considered\cite{PRE02016128,PRE04030902}.

The system evolves in time according to the following rules
\cite{NAP2007276}. The dynamics proceeds in parallel and considers a
discrete time step representing the fixed time scale $\tau $ of the
process. The reaction and diffusion rates are therefore converted
into probabilities. At each time step, the system is updated as
follows. Inside each network node with degree $k$, each infected
individual is cured and becomes an susceptible one with probability
$\mu_k \tau$. At the same time, each susceptible individual acquires
infection from any infected one with probability $1 - (1 - \beta
\tau )^{n_I }$, where $n_I$ is the total number of infected
individuals in the node. After all nodes have been updated for the
reactions, diffusion processes take place by allowing each
individual to move into a randomly chosen neighboring node with
probability $D_I \tau$ and $D_S \tau$, for infected and susceptible
individuals respectively, where $D_I(D_S)$ denotes the corresponding
diffusion constant. In our simulation, the parameters are $N=1000$,
$\beta=0.5$, $D_I=D_S=1.0$, $\tau =0.001$ if not otherwise
specified. Each plot is obtained via averaging over 20 independent
simulation runs.

\section{Simulation results}  \label{sec3}

\begin{figure}[h]
\centerline{\includegraphics*[width=0.436 \columnwidth]{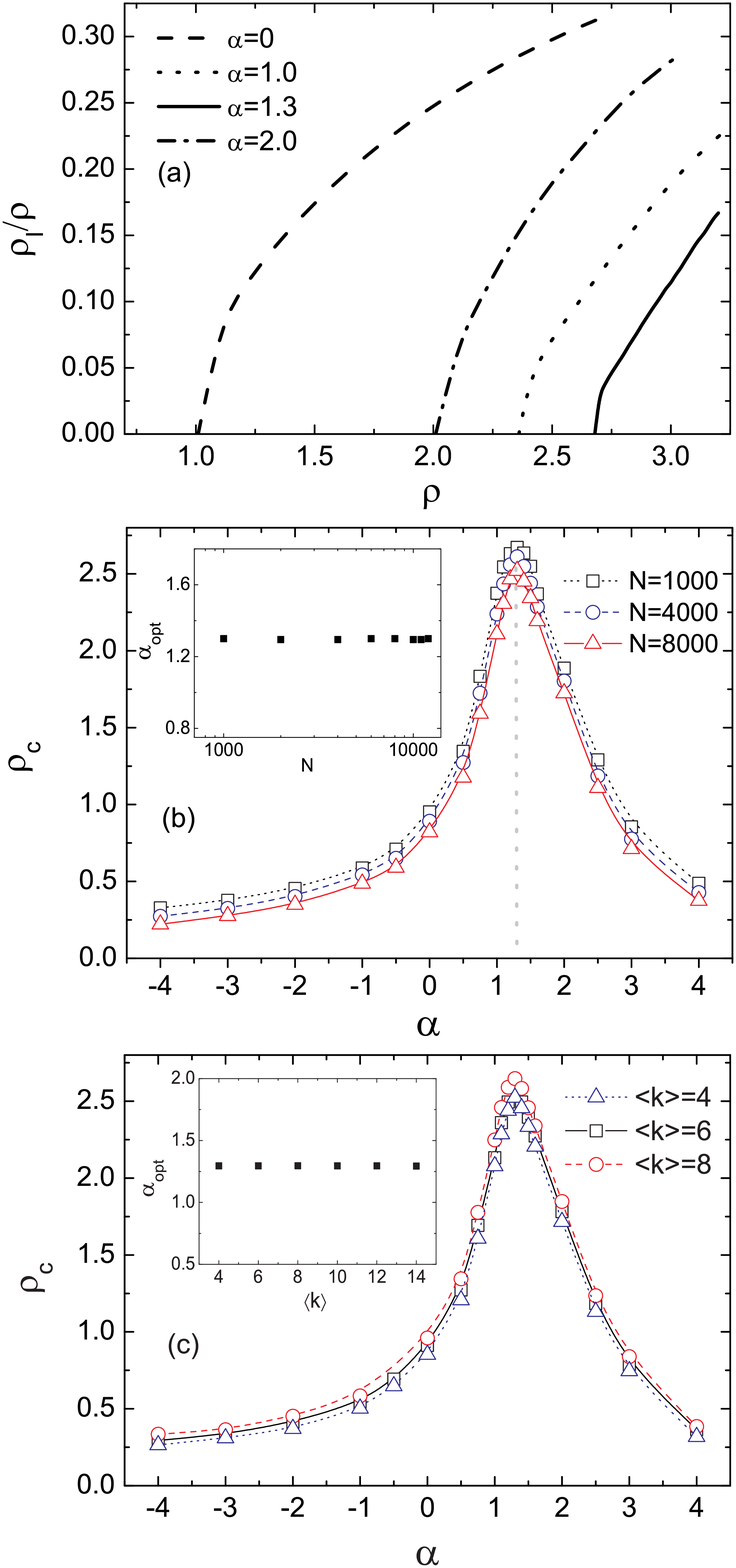}}
\caption{(Color online) (a) The proportion $\rho_I/\rho$ of infected
individuals as a function of $\rho$ for different $\alpha$ on
1000-node BA networks. (b) The epidemic threshold $\rho_c$ as a
function of $\alpha$ for different network sizes $N$. The maximal
threshold places $\alpha_{opt}\simeq 1.3$, which is indicated by
vertical dotted line. The inset shows $\alpha_{opt}$ as a function
of $N$. All the networks have the fixed average network degree
$\langle k\rangle =6$. (c)The epidemic threshold $\rho_c$ as a
function of $\alpha$ for different $\langle k\rangle$. The inset
shows $\alpha_{opt}$ as a function of $\langle k\rangle$. $N =4000$.
\label{figPhaseTransion}}
\end{figure}

Fig.\ref{figPhaseTransion}(a) shows how the proportion $\rho_I/\rho$
of infected individuals in the whole network increases with $\rho$,
where $\rho_I=\sum n_I/N$ denotes the density of infected
individuals in the whole network, for four different values of
$\alpha$. Clearly, the system  undergos a phase transition at a
certain threshold density $\rho_c$,  above which $\rho_I/\rho$
monotonically increases from zero. For $\rho<\rho_c$, the system
stays in a {\lq healthy\rq} state with $\rho_I=0$. Interestingly,
$\rho_c$ reaches a largest value for $\alpha$ =1.3, compared to
those for $\alpha$ =0, 1.0, 2.0.  This is demonstrated more clearly
in Fig.\ref{figPhaseTransion} (b), where $\rho_c$ are plotted as a
function of $\alpha$  for different network sizes $N$.  The distinct
peak locates at $\alpha_{opt} \simeq 1.3$, which is rather robust to
the change of network size $N$ as shown in the inset. In addition,
we have also investigated how this phenomenon depends on the average
network degree $\langle k\rangle$. As shown in
Fig.\ref{figPhaseTransion} (c), the optimal value $\alpha_{opt}$
also remains nearly constant with varying $\langle k\rangle$ from 4
to 14.

So far we have considered that all species diffuse with the same
rate. In the following, we will take into account different
diffusion rates for different species. For the sake of simplicity,
we assume that infected individuals diffuse with a fixed rate $D_I =
1$ and vary the diffusion rate of susceptible individuals $D_S$. The
epidemic threshold $\rho_c$ as a function of $\alpha$ is plotted  in
Fig.2 for $D_S =0, 0.005, 0.05, 0.5$, and $1.0 $.  Interestingly,
the bell-shape dependence of $\rho_c$ on $\alpha$ always exists for
nonzero $D_S$, with the peak located at nearly the same optimal
value $\alpha_{opt} \simeq1.3$. The height of this peak decreases
with $D_S$, and eventually $\rho_c$ is independent of $\alpha$ for
$D_S=0$.

\begin{figure}[h]
\centerline{\includegraphics*[width=0.55 \columnwidth]{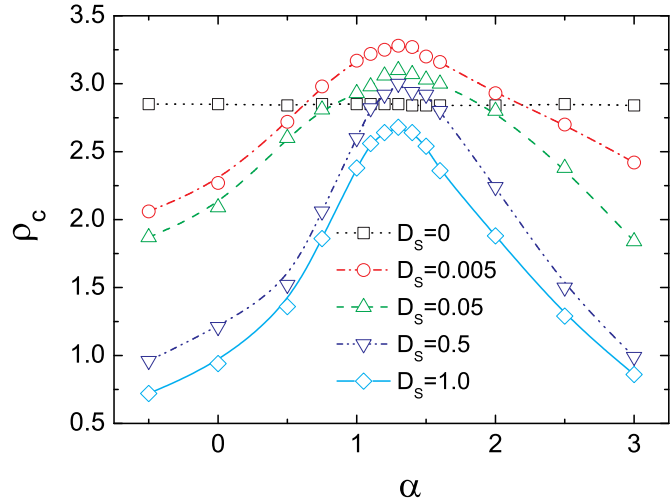}}
\caption{(Color online) The epidemic threshold $\rho_c$ as a
function of $\alpha$ for different diffusion rates $D_S$. All the
networks have the fixed $\langle k\rangle =6$, $N =1000$ and
$\gamma=3.0$. \label{fig2}}
\end{figure}

Fig. 3(a) shows that $\rho_c$ as a function of $\alpha$ for
different infection rates $\beta$. It can be found that the values
of $\beta$ do not influence the qualitative dependence of $\rho_c$
on $\alpha$, i.e., a maximum $\rho_c$ still shows up for the same
 optimal $\alpha$. Nevertheless, the maximum $\rho_c$ corresponding to $\alpha_{opt}$
do change with $\beta$. In addition, we have also considered how the
above findings depend on the network topology. To this end, we have
performed simulations on SF networks with different exponents
$\gamma$ and Erd\"{o}s-R\'{e}nyi (ER) random networks.  $\rho_c$ as
a function of $\alpha$ for different type of networks are shown in
Fig.3(b). It is found that there still exists an optimal value of
$\alpha$, leading to the maximal threshold. For SF networks, the
optimal value of $\alpha$ is always close to $1.3$, while for ER
networks, the $\rho_c \sim \alpha$ curve becomes not so sharp
indicating that $\rho_c$ is not sensitive to the change of $\alpha$.

\begin{figure}[h]
\centerline{\includegraphics*[width=0.55 \columnwidth]{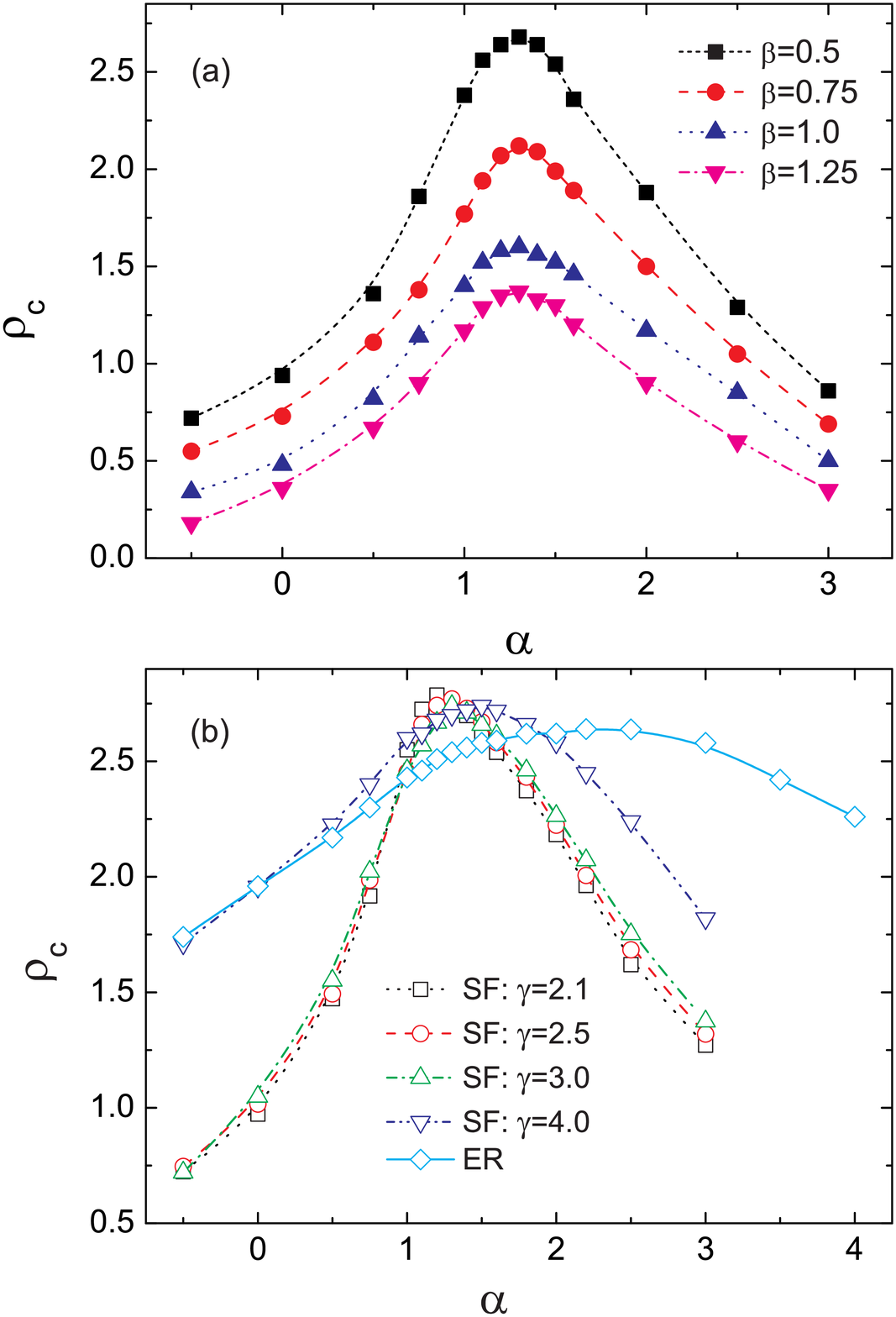}}
\caption{(Color online) The epidemic threshold $\rho_c$ as a
function of $\alpha$ for different infection rates $\beta$ (a) and
for different network topologies (b). For (b), $\beta=0.5$. All the
networks have the fixed $\langle k\rangle =6$ and $N =1000$.
\label{fig3}}
\end{figure}

\section{Mean field analysis} \label{sec4}

According to the stochastic simulation scheme, one may write down the following set of dynamic equations at a MF level,
\begin{subequations} \label{eqRateModel}
\begin{eqnarray}
  \frac{{\partial \rho _{I,k} }}
{{\partial t}} =  \rho _{I,k}(\beta \rho _{S,k}- \mu _k) + D_I
\left( {k\sum\limits_{k'} {p(k'\left| k \right.)} \frac{1}
{{k'}}\rho _{I,k'}  - \rho _{I,k} } \right)  \\
  \frac{{\partial \rho _{S,k} }}
{{\partial t}} = \rho _{I,k}(\mu _k   - \beta \rho _{S,k}) + D_S
\left( {k\sum\limits_{k'} {p(k'\left| k \right.)} \frac{1}
{{k'}}\rho _{S,k'}  - \rho _{S,k} } \right)
\end{eqnarray}
\end{subequations}
where $\rho_{I,k}$ and $\rho _{S,k} $ represent the average
densities  of infected and susceptible individuals, respectively, in
the nodes with degree $k$. The first term in the right hand side of
Eq.(3a) accounts for the change of infected individuals due to the
reaction (infection and recovery) processes, and the second term
accounts for the diffusion of infected individuals into and out of
those nodes with degree $k$. Eq.(3b) can be interpreted in a similar
manner. $ p(k'\left| k \right.)$ represents the conditional
probability that a node of degree $k$ is connected to a node of
degree $k'$, which equals to $ k'p(k')/\langle k'\rangle $
\cite{PRL01258701,Dor2003}  for BA networks.

One notes that a thorough analysis of Eqs.(3) is not easy. For sake of simplicity, here we only consider the case $D_I=D_S=1$.
Then,  substituting this into
Eqs.(\ref{eqRateModel})  and using $\rho_I=\sum_k p(k) \rho_{I,k}$,
one obtains
\begin{subequations} \label{eqRate}
\begin{eqnarray}
\frac{{\partial \rho _{I,k} }} {{\partial t}} = \rho _{I,k} (\beta
\rho _{S,k} - \mu _k ) + \frac{k}
{{\langle k\rangle }}\rho _I  - \rho _{I,k}   \\
  \frac{{\partial \rho _{S,k} }}
{{\partial t}} = \rho _{I,k} ( \mu _k  - \beta \rho _{S,k} ) +
\frac{k} {{\langle k\rangle }}\rho _S  - \rho _{S,k}
\end{eqnarray}
\end{subequations}

 For $\alpha=0$ and thus $\mu _k=1$, it is already shown that
$ \rho _c = \frac{\mu } {\beta }\frac{{\langle
k\rangle ^2 }} {{\langle k^2 \rangle }}$ \cite{NAP2007276}. But for
$\alpha\neq 0$, it is hard to get the explicit expression of $\rho_c$  from Eqs.(\ref{eqRate}) directly. Clearly, Eqs.(\ref{eqRate}) admit a
steady state, which solves $\partial \rho _{I,k} / \partial t=\partial \rho _{S,k} / \partial t=0$,
\begin{equation} \label{stationarySolve}
\rho _{^{I,k} }^*  = 0, {\kern 10pt} \rho _{_{S,k} }^*  = \frac{k}
{{\langle k\rangle }}\rho
\end{equation}
which physically corresponds to the disease-free state. Intuitively,
this healthy state will lose stability at the critical density
$\rho_c$, above which the steady state value of $\rho_{I,K}$ cannot
be 0 any more. Therefore, one can alternatively perform linear
stability analysis of $(\rho_{^{I,k} }^* ,\rho_{^{S,k} }^* ) $ to
get $\rho_c$.  Following standard procedures, one can readily obtain
the Jacobian matrix and calculate the eigenvalues $\{\lambda\}$. The
healthy state will lose stability when $\lambda_{max}$, the largest
value of the real part of the eigenvalues, passes through zero from
below. Note that explicit expression for $\lambda_{max}$ is not
available, but numerical calculation of it is easy.

\begin{figure}[h]
\centerline{\includegraphics*[width=0.55 \columnwidth]{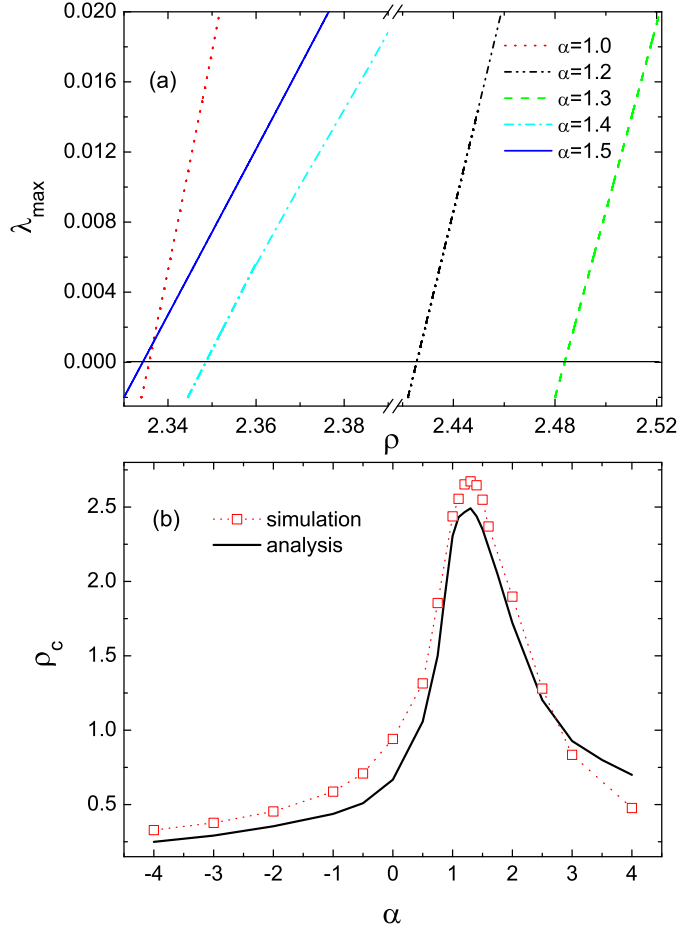}}
\caption{(Color online) (a) and (b) correspond to the dependence of
the largest eigenvalues $\lambda_{max} $ on $\rho$ for different
$\alpha$ and the dependence of $\rho_c$ on $\alpha$ respectively,
both on a synthesized 1000-node BA network and $\langle k\rangle
=6$. \label{figEigenvalue}}
\end{figure}

Fig. \ref{figEigenvalue} (a) plots $\lambda_{max}$  as a function of
$\rho$ for several values of  $\alpha$. The value of $\rho$ where
the $\lambda_{max}=0$ corresponds to $\rho_c$. As expected, $\rho_c$
is the largest for $\alpha=1.3$  compared to those for other
$\alpha$. Fig. \ref{figEigenvalue} (b) presents $\rho_c$ as a
function of $\alpha$ obtained from simulations (symbols) and MF
analysis (solid line). Apparently, the MF results are in rather good
agreements with the simulation ones in Fig.1.

\begin{figure}[h]
\centerline{\includegraphics*[width=0.55 \columnwidth]{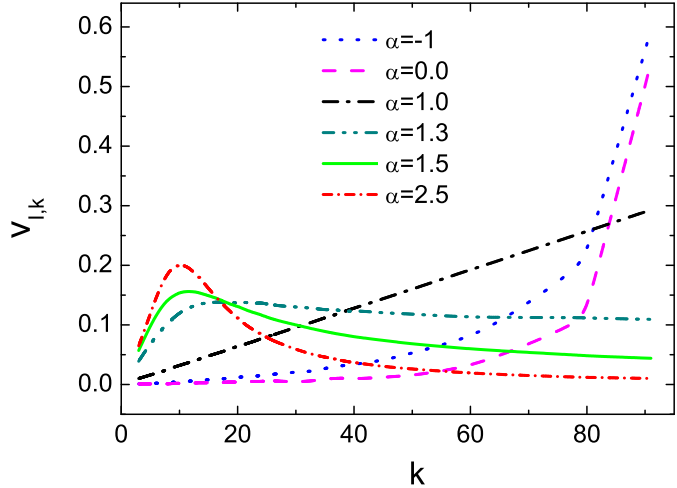}}
\caption{(Color online) The eigenvectors $\textbf{v}_{I,k}$
corresponding to the dominant eigenvalue $\lambda_{max}=0$, as a
function of $k$ for different $\alpha$. Other parameters are the
same as in Fig. \ref{figEigenvalue}. \label{figEigenvector}}
\end{figure}

To get more insights into how the epidemic explosion takes place for
different $\alpha$, we turn to the eigenvector $\textbf{v}=\{
(\text{v}_{I,k},\text{v}_{S,k})_{k=1,\dots} \}$ corresponding to
$\lambda_{max}$ at the onset of the phase transition, i.e.,
$\rho=\rho_c$. The element $\text{v}_{I,k}$ of this vector measures
the relative amplitude of the fluctuation away from $\rho^*_{I,k}=0$
for nodes with given degree $k$. Therefore, the dependence of
$\text{v}_{I,k}$ on $k$ qualitatively tells us how the epidemic
explosion grows from the healthy state. In Fig.
\ref{figEigenvector}, we  depict the eigenvectors $\textbf{v}_{I,k}$
as a function of $k$ for different $\alpha$. Interestingly, it can
be observed  that epidemic explosion starts from large-degree nodes
for $\alpha$ less than $\alpha_{opt}$, as shown by the {\lq\lq
dotted\rq\rq}, {\lq\lq dashed\rq\rq} and {\lq\lq dash dotted\rq\rq}
lines in Fig.\ref{figEigenvector}, while it is from small-degree
nodes for $\alpha$ larger than $\alpha_{opt}$, as shown by {\lq\lq
solid\rq\rq} and {\lq\lq short dash dotted\rq\rq} lines. For $\alpha
\simeq \alpha_{opt}$, $\text{v}_{I,k}$ is not that sensitive on $k$,
indicating a relatively homogenous epidemic explosion.

\begin{figure}[h]
\centerline{\includegraphics*[width=0.55 \columnwidth]{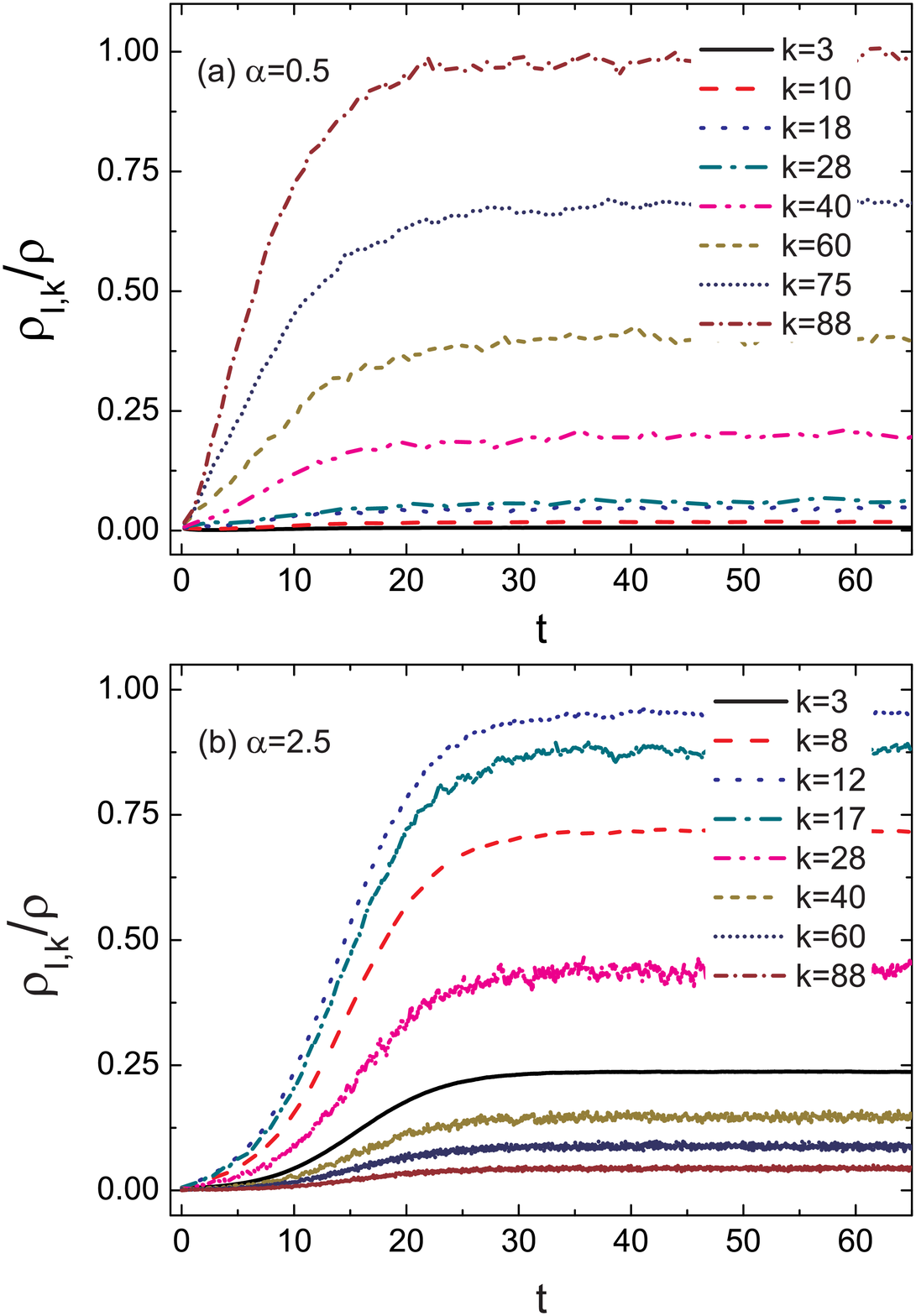}}
\caption{(Color online) Time evolution $\rho_{I,k}/ \rho$ of the
average density of infected individuals in the nodes with different
degrees for BA model with
 $N =1000$ and $\langle k\rangle =6$. (a) $\alpha=0.5$, (b) $\alpha=2.5$. \label{fig6}}
\end{figure}

To reveal the underlying mechanism of the epidemic spreading for
different $\alpha$ in more detail, we illustrate the time evolution
of $\rho_{I,k}/ \rho$, the average density of infected individuals
in the nodes with degree $k$ in Fig. \ref{fig6} for two particular
values of $\alpha$, one ($\alpha=0.5$) less than $\alpha_{opt}$ and
the other ($\alpha=2.5$) larger than it. This can give us more
detailed information about how the epidemic outbreak takes place on
nodes with different degree $k$ . We find that, for $\alpha=0.5$,
the disease starts to spread from large-degree nodes, such as
$k$=88, 75 and 60, as shown by the  top three lines in Fig. 6(a);
while for $\alpha=2.5$, the spreading starts from those nodes with
relatively small degree, such as $k$=12, 17 and 8, as shown in Fig.
6(b). These phenomena indicate that there indeed exist two different
epidemic explosion routes for $\alpha$ being less or larger than
$\alpha_{opt}$, which are consistent with the analysis associated
with the eigenvectors as shown in Fig.5.

The above different pathways regarding small or large $\alpha$ may
be illustrated qualitatively in the following way. Consider the
individuals in a given node are infected at the beginning. These
patients will diffuse to neighboring nodes through the links.
Certainly, nodes with larger degrees will have more chances to
accept these patients. To efficiently suppress the epidemic
explosion, the curing rates in such large-degree nodes should be
relatively large to compensate these incoming patients via
diffusion. Therefore, it is reasonable that $\mu_k$ should be an
increasing function of $k$ to maintain an effective epidemic
control. Intuitively, one may imagine that the most efficient way is
to keep linear dependence of $\mu_k$ on $k$, i.e., $\alpha=1$ in our
strategy, considering that every incoming patient via diffusion can
be cured on time. However, this is not exactly the case because the
reactions inside a node involve nonlinear autocatalytic processes,
which makes  $\alpha_{opt}$ larger than 1 (Unfortunately, why
$\alpha_{opt}$ is so robust to be about 1.3 is still open to us). If
$\alpha$ is too large, which means that the medical resources are
biased to large-degree nodes, the patients in small-degree nodes
cannot be cured on time. In this case, disease will start to spread
from those small-degree nodes. In the contrast case, the disease
will start more abruptly from those large-degree nodes since the
curing rates there are too smaller than required. These scenario are
in agreement with the picture shown in Fig. \ref{figEigenvector} and
\ref{fig6}.

\section{Discussion and Conclusions} \label{sec7}

One should note that the $\alpha$ value can not be arbitrary for the
real world, if we accept the power-law dependence. Following the
recipe of Eq.2, for a scale free network with minimum, mean and
maximum degree respectively of 2, 5 and 100, the recovery rate will
range from 0.4 to 20 in the simplest case of linear dependency
($\alpha=1$). This large difference is to some extent not
reasonable, which  implies that  the optimal control with
$\alpha=1.3$ is hard to be realized practically. Nevertheless, as a
model study, we can just change $\alpha$ as we want to see what we
can find. If, for instance, we tune $\alpha$ to a reasonable
non-zero value, say $\alpha=0.5$, the ratio of the maximal and
minimal $\mu$ would be about 10 for a network with $k$ ranging from
1 to 100, which can also lead to a much better epidemic control
($\rho_c=1.37$) than previous case of $\alpha=0$ ($\rho_c=0.95$).
Therefore, our work has indeed provided an efficient strategy to
suppress the epidemic explosion.

In summary, we have studied a variant of  $SIS$ model defined on
scale-free metapopulation networks, wherein the curing rate in a
node with degree $k$ is proportional to $k^\alpha$. By detailed
numerical simulations, we show that the epidemic threshold reaches a
maximum value when $\alpha$ is tuned to be $\alpha_{opt}\simeq 1.3$,
which corresponds to an optimal control strategy to suppress
epidemic explosion and is robust to the change of network size or
average degree.  We have also performed a mean field analysis to
further elucidate this strategy and unravel the distinct pathways to
epidemic spreading for $\alpha$ larger or less than $\alpha_{opt}$.
Our findings suggest that a proper allocation of medical resources
can best suppress the epidemic explosion, which could be of great
importance in practical epidemic control.

\begin{acknowledgments}
This work was supported by the National Natural Science Foundation
of China (Grant Nos.21125313, 20933006 and 91027012). C.S.S was also
supported by the Key Scientific Research Fund of Anhui Provincial
Education Department (Grant No. KJ2012A189).
\end{acknowledgments}


\bibliographystyle{apsrev}

\begin{thebibliography}{49}
\expandafter\ifx\csname
natexlab\endcsname\relax\def\natexlab#1{#1}\fi
\expandafter\ifx\csname bibnamefont\endcsname\relax
  \def\bibnamefont#1{#1}\fi
\expandafter\ifx\csname bibfnamefont\endcsname\relax
  \def\bibfnamefont#1{#1}\fi
\expandafter\ifx\csname citenamefont\endcsname\relax
  \def\citenamefont#1{#1}\fi
\expandafter\ifx\csname url\endcsname\relax
  \def\url#1{\texttt{#1}}\fi
\expandafter\ifx\csname urlprefix\endcsname\relax\def\urlprefix{URL
}\fi \providecommand{\bibinfo}[2]{#2}
\providecommand{\eprint}[2][]{\url{#2}}

\bibitem[{\citenamefont{Albert and Barab\'{a}si}(2002)}]{RMP02000047}
\bibinfo{author}{\bibfnamefont{R.}~\bibnamefont{Albert}} \bibnamefont{and}
  \bibinfo{author}{\bibfnamefont{A.-L.} \bibnamefont{Barab\'{a}si}},
  \bibinfo{journal}{Rev. Mod. Phys.} \textbf{\bibinfo{volume}{74}},
  \bibinfo{pages}{47} (\bibinfo{year}{2002}).

\bibitem[{\citenamefont{Dorogovtsev and Mendes}(2002)}]{AIP02001079}
\bibinfo{author}{\bibfnamefont{S.~N.} \bibnamefont{Dorogovtsev}}
  \bibnamefont{and} \bibinfo{author}{\bibfnamefont{J.~F.~F.}
  \bibnamefont{Mendes}}, \bibinfo{journal}{Adv. Phys.}
  \textbf{\bibinfo{volume}{51}}, \bibinfo{pages}{1079} (\bibinfo{year}{2002}).

\bibitem[{\citenamefont{Newman}(2003)}]{SIR03000167}
\bibinfo{author}{\bibfnamefont{M.~E.~J.} \bibnamefont{Newman}},
  \bibinfo{journal}{SIAM Review} \textbf{\bibinfo{volume}{45}},
  \bibinfo{pages}{167} (\bibinfo{year}{2003}).

\bibitem[{\citenamefont{Boccaletti et~al.}(2006)\citenamefont{Boccaletti,
  Latora, Moreno, Chavez, and Hwang}}]{PRP06000175}
\bibinfo{author}{\bibfnamefont{S.}~\bibnamefont{Boccaletti}},
  \bibinfo{author}{\bibfnamefont{V.}~\bibnamefont{Latora}},
  \bibinfo{author}{\bibfnamefont{Y.}~\bibnamefont{Moreno}},
  \bibinfo{author}{\bibfnamefont{M.}~\bibnamefont{Chavez}}, \bibnamefont{and}
  \bibinfo{author}{\bibfnamefont{D.-U.} \bibnamefont{Hwang}},
  \bibinfo{journal}{Phys. Rep.} \textbf{\bibinfo{volume}{424}},
  \bibinfo{pages}{175} (\bibinfo{year}{2006}).

\bibitem[{\citenamefont{Arenas et~al.}(2008)\citenamefont{Arenas,
  D\'iaz-Guilera, Kurths, Moreno, and Zhou}}]{PRP08000093}
\bibinfo{author}{\bibfnamefont{A.}~\bibnamefont{Arenas}},
  \bibinfo{author}{\bibfnamefont{A.}~\bibnamefont{D\'iaz-Guilera}},
  \bibinfo{author}{\bibfnamefont{J.}~\bibnamefont{Kurths}},
  \bibinfo{author}{\bibfnamefont{Y.}~\bibnamefont{Moreno}}, \bibnamefont{and}
  \bibinfo{author}{\bibfnamefont{C.}~\bibnamefont{Zhou}},
  \bibinfo{journal}{Phys. Rep.} \textbf{\bibinfo{volume}{469}},
  \bibinfo{pages}{93} (\bibinfo{year}{2008}).

\bibitem[{\citenamefont{Dorogovtsev et~al.}(2008)\citenamefont{Dorogovtsev,
  Goltsev, and Mendes}}]{RMP08001275}
\bibinfo{author}{\bibfnamefont{S.~N.} \bibnamefont{Dorogovtsev}},
  \bibinfo{author}{\bibfnamefont{A.~V.} \bibnamefont{Goltsev}},
  \bibnamefont{and} \bibinfo{author}{\bibfnamefont{J.~F.~F.}
  \bibnamefont{Mendes}}, \bibinfo{journal}{Rev. Mod. Phys.}
  \textbf{\bibinfo{volume}{80}}, \bibinfo{pages}{1275} (\bibinfo{year}{2008}).

\bibitem[{\citenamefont{Pastor-Satorras and
  Vespignani}(2001{\natexlab{a}})}]{PRL01003200}
\bibinfo{author}{\bibfnamefont{R.}~\bibnamefont{Pastor-Satorras}}
  \bibnamefont{and}
  \bibinfo{author}{\bibfnamefont{A.}~\bibnamefont{Vespignani}},
  \bibinfo{journal}{Phys. Rev. Lett.} \textbf{\bibinfo{volume}{86}},
  \bibinfo{pages}{3200} (\bibinfo{year}{2001}{\natexlab{a}}).

\bibitem[{\citenamefont{Pastor-Satorras and
  Vespignani}(2001{\natexlab{b}})}]{PRE01066117}
\bibinfo{author}{\bibfnamefont{R.}~\bibnamefont{Pastor-Satorras}}
  \bibnamefont{and}
  \bibinfo{author}{\bibfnamefont{A.}~\bibnamefont{Vespignani}},
  \bibinfo{journal}{Phys. Rev. E} \textbf{\bibinfo{volume}{63}},
  \bibinfo{pages}{066117} (\bibinfo{year}{2001}{\natexlab{b}}).

\bibitem[{\citenamefont{Bogu\~n\'a and
  Pastor-Satorras}(2002)}]{PhysRevE.66.047104}
\bibinfo{author}{\bibfnamefont{M.}~\bibnamefont{Bogu\~n\'a}} \bibnamefont{and}
  \bibinfo{author}{\bibfnamefont{R.}~\bibnamefont{Pastor-Satorras}},
  \bibinfo{journal}{Phys. Rev. E} \textbf{\bibinfo{volume}{66}},
  \bibinfo{pages}{047104} (\bibinfo{year}{2002}).

\bibitem[{\citenamefont{Bogu\~n\'a et~al.}(2003)\citenamefont{Bogu\~n\'a,
  Pastor-Satorras, and Vespignani}}]{PRL03028701}
\bibinfo{author}{\bibfnamefont{M.}~\bibnamefont{Bogu\~n\'a}},
  \bibinfo{author}{\bibfnamefont{R.}~\bibnamefont{Pastor-Satorras}},
  \bibnamefont{and}
  \bibinfo{author}{\bibfnamefont{A.}~\bibnamefont{Vespignani}},
  \bibinfo{journal}{Phys. Rev. Lett.} \textbf{\bibinfo{volume}{90}},
  \bibinfo{pages}{028701} (\bibinfo{year}{2003}).

\bibitem[{\citenamefont{Barth\'elemy et~al.}(2004)\citenamefont{Barth\'elemy,
  Barrat, Pastor-Satorras, and Vespignani}}]{PRL04178701}
\bibinfo{author}{\bibfnamefont{M.}~\bibnamefont{Barth\'elemy}},
  \bibinfo{author}{\bibfnamefont{A.}~\bibnamefont{Barrat}},
  \bibinfo{author}{\bibfnamefont{R.}~\bibnamefont{Pastor-Satorras}},
  \bibnamefont{and}
  \bibinfo{author}{\bibfnamefont{A.}~\bibnamefont{Vespignani}},
  \bibinfo{journal}{Phys. Rev. Lett.} \textbf{\bibinfo{volume}{92}},
  \bibinfo{pages}{178701} (\bibinfo{year}{2004}).

\bibitem[{\citenamefont{May and Lloyd}(2001)}]{PRE01066112}
\bibinfo{author}{\bibfnamefont{R.~M.} \bibnamefont{May}} \bibnamefont{and}
  \bibinfo{author}{\bibfnamefont{A.~L.} \bibnamefont{Lloyd}},
  \bibinfo{journal}{Phys. Rev. E} \textbf{\bibinfo{volume}{64}},
  \bibinfo{pages}{066112} (\bibinfo{year}{2001}).

\bibitem[{\citenamefont{Lloyd and May}(2001)}]{SCI01001316}
\bibinfo{author}{\bibfnamefont{A.~L.} \bibnamefont{Lloyd}} \bibnamefont{and}
  \bibinfo{author}{\bibfnamefont{R.~M.} \bibnamefont{May}},
  \bibinfo{journal}{Science} \textbf{\bibinfo{volume}{292}},
  \bibinfo{pages}{1316} (\bibinfo{year}{2001}).

\bibitem[{\citenamefont{Moreno et~al.}(2002)\citenamefont{Moreno,
  Pastor-Satorras, and Vespignani}}]{EPJB02000521}
\bibinfo{author}{\bibfnamefont{Y.}~\bibnamefont{Moreno}},
  \bibinfo{author}{\bibfnamefont{R.}~\bibnamefont{Pastor-Satorras}},
  \bibnamefont{and}
  \bibinfo{author}{\bibfnamefont{A.}~\bibnamefont{Vespignani}},
  \bibinfo{journal}{Eur. Phys. J. B} \textbf{\bibinfo{volume}{26}},
  \bibinfo{pages}{521} (\bibinfo{year}{2002}).

\bibitem[{\citenamefont{Moreno et~al.}(2003)\citenamefont{Moreno, G\'omez, and
  Pacheco}}]{PRE03035103}
\bibinfo{author}{\bibfnamefont{Y.}~\bibnamefont{Moreno}},
  \bibinfo{author}{\bibfnamefont{J.~B.} \bibnamefont{G\'omez}},
  \bibnamefont{and} \bibinfo{author}{\bibfnamefont{A.~F.}
  \bibnamefont{Pacheco}}, \bibinfo{journal}{Phys. Rev. E}
  \textbf{\bibinfo{volume}{68}}, \bibinfo{pages}{035103}
  (\bibinfo{year}{2003}).

\bibitem[{\citenamefont{Joo and Lebowitz}(2004)}]{PRE04066105}
\bibinfo{author}{\bibfnamefont{J.}~\bibnamefont{Joo}} \bibnamefont{and}
  \bibinfo{author}{\bibfnamefont{J.~L.} \bibnamefont{Lebowitz}},
  \bibinfo{journal}{Phys. Rev. E} \textbf{\bibinfo{volume}{69}},
  \bibinfo{pages}{066105} (\bibinfo{year}{2004}).

\bibitem[{\citenamefont{Olinky and Stone}(2004)}]{PRE04030902}
\bibinfo{author}{\bibfnamefont{R.}~\bibnamefont{Olinky}} \bibnamefont{and}
  \bibinfo{author}{\bibfnamefont{L.}~\bibnamefont{Stone}},
  \bibinfo{journal}{Phys. Rev. E} \textbf{\bibinfo{volume}{70}},
  \bibinfo{pages}{030902(R)} (\bibinfo{year}{2004}).

\bibitem[{\citenamefont{Liu and Hu}(2005)}]{EPL05000315}
\bibinfo{author}{\bibfnamefont{Z.}~\bibnamefont{Liu}} \bibnamefont{and}
  \bibinfo{author}{\bibfnamefont{B.}~\bibnamefont{Hu}},
  \bibinfo{journal}{Europhys. Lett.} \textbf{\bibinfo{volume}{72}},
  \bibinfo{pages}{315} (\bibinfo{year}{2005}).

\bibitem[{\citenamefont{Yan et~al.}(2007)\citenamefont{Yan, Fu, Ren, and
  Wang}}]{PRE07016108}
\bibinfo{author}{\bibfnamefont{G.}~\bibnamefont{Yan}},
  \bibinfo{author}{\bibfnamefont{Z.-Q.} \bibnamefont{Fu}},
  \bibinfo{author}{\bibfnamefont{J.}~\bibnamefont{Ren}}, \bibnamefont{and}
  \bibinfo{author}{\bibfnamefont{W.-X.} \bibnamefont{Wang}},
  \bibinfo{journal}{Phys. Rev. E} \textbf{\bibinfo{volume}{75}},
  \bibinfo{pages}{016108} (\bibinfo{year}{2007}).

\bibitem[{\citenamefont{Castellano and Pastor-Satorras}(2010)}]{PRL10218701}
\bibinfo{author}{\bibfnamefont{C.}~\bibnamefont{Castellano}} \bibnamefont{and}
  \bibinfo{author}{\bibfnamefont{R.}~\bibnamefont{Pastor-Satorras}},
  \bibinfo{journal}{Phys. Rev. Lett.} \textbf{\bibinfo{volume}{105}},
  \bibinfo{pages}{218701} (\bibinfo{year}{2010}).

\bibitem[{\citenamefont{Kitsak et~al.}(2010)\citenamefont{Kitsak, Gallos,
  Havlin, Liljeros, Muchnik, Stanley, and Makse}}]{NAP2010888}
\bibinfo{author}{\bibfnamefont{M.}~\bibnamefont{Kitsak}},
  \bibinfo{author}{\bibfnamefont{L.~K.} \bibnamefont{Gallos}},
  \bibinfo{author}{\bibfnamefont{S.}~\bibnamefont{Havlin}},
  \bibinfo{author}{\bibfnamefont{F.}~\bibnamefont{Liljeros}},
  \bibinfo{author}{\bibfnamefont{L.}~\bibnamefont{Muchnik}},
  \bibinfo{author}{\bibfnamefont{H.~E.} \bibnamefont{Stanley}},
  \bibnamefont{and} \bibinfo{author}{\bibfnamefont{H.~A.} \bibnamefont{Makse}},
  \bibinfo{journal}{Nature Phys.} \textbf{\bibinfo{volume}{8}},
  \bibinfo{pages}{888} (\bibinfo{year}{2010}).

\bibitem[{\citenamefont{Masuda}(2010)}]{NJP10093009}
\bibinfo{author}{\bibfnamefont{N.}~\bibnamefont{Masuda}}, \bibinfo{journal}{New
  J. Phys.} \textbf{\bibinfo{volume}{12}}, \bibinfo{pages}{093009}
  (\bibinfo{year}{2010}).

\bibitem[{\citenamefont{Neri et~al.}(2011)\citenamefont{Neri, Bates,
  F\"{u}chtbauer, P\'{e}rez-Reche, Taraskin, Otten, Bailey, and
  Gilligan}}]{PLoS111002174}
\bibinfo{author}{\bibfnamefont{F.~M.} \bibnamefont{Neri}},
  \bibinfo{author}{\bibfnamefont{A.}~\bibnamefont{Bates}},
  \bibinfo{author}{\bibfnamefont{W.~S.} \bibnamefont{F\"{u}chtbauer}},
  \bibinfo{author}{\bibfnamefont{F.~J.} \bibnamefont{P\'{e}rez-Reche}},
  \bibinfo{author}{\bibfnamefont{S.~N.} \bibnamefont{Taraskin}},
  \bibinfo{author}{\bibfnamefont{W.}~\bibnamefont{Otten}},
  \bibinfo{author}{\bibfnamefont{D.~J.} \bibnamefont{Bailey}},
  \bibnamefont{and} \bibinfo{author}{\bibfnamefont{C.~A.}
  \bibnamefont{Gilligan}}, \bibinfo{journal}{PLoS Comput. Biol.}
  \textbf{\bibinfo{volume}{7}}, \bibinfo{pages}{e1002174}
  (\bibinfo{year}{2011}).

\bibitem[{\citenamefont{Belik et~al.}(2011)\citenamefont{Belik, Geisel, and
  Brockmann}}]{PhysRevX.1.011001}
\bibinfo{author}{\bibfnamefont{V.}~\bibnamefont{Belik}},
  \bibinfo{author}{\bibfnamefont{T.}~\bibnamefont{Geisel}}, \bibnamefont{and}
  \bibinfo{author}{\bibfnamefont{D.}~\bibnamefont{Brockmann}},
  \bibinfo{journal}{Phys. Rev. X} \textbf{\bibinfo{volume}{1}},
  \bibinfo{pages}{011001} (\bibinfo{year}{2011}).

\bibitem[{\citenamefont{Colizza et~al.}(2007)\citenamefont{Colizza,
  Pastor-Satorras, and Vespignani}}]{NAP2007276}
\bibinfo{author}{\bibfnamefont{V.}~\bibnamefont{Colizza}},
  \bibinfo{author}{\bibfnamefont{R.}~\bibnamefont{Pastor-Satorras}},
  \bibnamefont{and}
  \bibinfo{author}{\bibfnamefont{A.}~\bibnamefont{Vespignani}},
  \bibinfo{journal}{Nature Phys.} \textbf{\bibinfo{volume}{3}},
  \bibinfo{pages}{276} (\bibinfo{year}{2007}).

\bibitem[{\citenamefont{Colizza and Vespignani}(2007)}]{PRL07148701}
\bibinfo{author}{\bibfnamefont{V.}~\bibnamefont{Colizza}} \bibnamefont{and}
  \bibinfo{author}{\bibfnamefont{A.}~\bibnamefont{Vespignani}},
  \bibinfo{journal}{Phys. Rev. Lett.} \textbf{\bibinfo{volume}{99}},
  \bibinfo{pages}{148701} (\bibinfo{year}{2007}).

\bibitem[{\citenamefont{Colizza and Vespignani}(2008)}]{JTB2008450}
\bibinfo{author}{\bibfnamefont{V.}~\bibnamefont{Colizza}} \bibnamefont{and}
  \bibinfo{author}{\bibfnamefont{A.}~\bibnamefont{Vespignani}},
  \bibinfo{journal}{J. Theor. Biol.} \textbf{\bibinfo{volume}{251}},
  \bibinfo{pages}{450} (\bibinfo{year}{2008}).

\bibitem[{\citenamefont{Gautreau et~al.}(2008)\citenamefont{Gautreau, Barrat,
  and Barth\'{e}lemy}}]{JTB2008509}
\bibinfo{author}{\bibfnamefont{A.}~\bibnamefont{Gautreau}},
  \bibinfo{author}{\bibfnamefont{A.}~\bibnamefont{Barrat}}, \bibnamefont{and}
  \bibinfo{author}{\bibfnamefont{M.}~\bibnamefont{Barth\'{e}lemy}},
  \bibinfo{journal}{J. Theor. Biol.} \textbf{\bibinfo{volume}{251}},
  \bibinfo{pages}{509} (\bibinfo{year}{2008}).

\bibitem[{\citenamefont{Baronchelli et~al.}(2008)\citenamefont{Baronchelli,
  Catanzaro, and Pastor-Satorras}}]{PRE08016111}
\bibinfo{author}{\bibfnamefont{A.}~\bibnamefont{Baronchelli}},
  \bibinfo{author}{\bibfnamefont{M.}~\bibnamefont{Catanzaro}},
  \bibnamefont{and}
  \bibinfo{author}{\bibfnamefont{R.}~\bibnamefont{Pastor-Satorras}},
  \bibinfo{journal}{Phys. Rev. E} \textbf{\bibinfo{volume}{78}},
  \bibinfo{pages}{016111} (\bibinfo{year}{2008}).

\bibitem[{\citenamefont{Wang et~al.}(2009)\citenamefont{Wang, Gonz\'{a}lez,
  Hidalgo, and Barab\'{a}si}}]{SCI20091071}
\bibinfo{author}{\bibfnamefont{P.}~\bibnamefont{Wang}},
  \bibinfo{author}{\bibfnamefont{M.~C.} \bibnamefont{Gonz\'{a}lez}},
  \bibinfo{author}{\bibfnamefont{C.~A.} \bibnamefont{Hidalgo}},
  \bibnamefont{and} \bibinfo{author}{\bibfnamefont{A.-L.}
  \bibnamefont{Barab\'{a}si}}, \bibinfo{journal}{Science}
  \textbf{\bibinfo{volume}{324}}, \bibinfo{pages}{1071} (\bibinfo{year}{2009}).

\bibitem[{\citenamefont{Kondo and Miura}(2010)}]{SCI20101616}
\bibinfo{author}{\bibfnamefont{S.}~\bibnamefont{Kondo}} \bibnamefont{and}
  \bibinfo{author}{\bibfnamefont{T.}~\bibnamefont{Miura}},
  \bibinfo{journal}{Science} \textbf{\bibinfo{volume}{329}},
  \bibinfo{pages}{1616} (\bibinfo{year}{2010}).

\bibitem[{\citenamefont{Nakamasu et~al.}(2009)\citenamefont{Nakamasu,
  Takahashi, Kanbe, and Kondo}}]{PNAS098429}
\bibinfo{author}{\bibfnamefont{A.}~\bibnamefont{Nakamasu}},
  \bibinfo{author}{\bibfnamefont{G.}~\bibnamefont{Takahashi}},
  \bibinfo{author}{\bibfnamefont{A.}~\bibnamefont{Kanbe}}, \bibnamefont{and}
  \bibinfo{author}{\bibfnamefont{S.}~\bibnamefont{Kondo}},
  \bibinfo{journal}{Proc. Natl. Acad. Sci. U.S.A.}
  \textbf{\bibinfo{volume}{106}}, \bibinfo{pages}{8429} (\bibinfo{year}{2009}).

\bibitem[{\citenamefont{Lizana et~al.}(2009)\citenamefont{Lizana, Konkoli,
  Bauer, Jesorka, and Orwar}}]{ARPC2009449}
\bibinfo{author}{\bibfnamefont{L.}~\bibnamefont{Lizana}},
  \bibinfo{author}{\bibfnamefont{Z.}~\bibnamefont{Konkoli}},
  \bibinfo{author}{\bibfnamefont{B.}~\bibnamefont{Bauer}},
  \bibinfo{author}{\bibfnamefont{A.}~\bibnamefont{Jesorka}}, \bibnamefont{and}
  \bibinfo{author}{\bibfnamefont{O.}~\bibnamefont{Orwar}},
  \bibinfo{journal}{Annu. Rev. Phys. Chem.} \textbf{\bibinfo{volume}{60}},
  \bibinfo{pages}{449} (\bibinfo{year}{2009}).

\bibitem[{\citenamefont{Grzybowski}(2009)}]{Grz2009}
\bibinfo{author}{\bibfnamefont{B.~A.} \bibnamefont{Grzybowski}},
  \emph{\bibinfo{title}{Chemistry in Motion: Reaction-Diffusion Systems for
  Micro- and Nanotechnology}} (\bibinfo{publisher}{Wiley, Chichester},
  \bibinfo{address}{West Sussex}, \bibinfo{year}{2009}).

\bibitem[{\citenamefont{Xuan et~al.}(2010)\citenamefont{Xuan, Du, Wu, and
  Chen}}]{PRE10046116}
\bibinfo{author}{\bibfnamefont{Q.}~\bibnamefont{Xuan}},
  \bibinfo{author}{\bibfnamefont{F.}~\bibnamefont{Du}},
  \bibinfo{author}{\bibfnamefont{T.-J.} \bibnamefont{Wu}}, \bibnamefont{and}
  \bibinfo{author}{\bibfnamefont{G.}~\bibnamefont{Chen}},
  \bibinfo{journal}{Phys. Rev. E} \textbf{\bibinfo{volume}{82}},
  \bibinfo{pages}{046116} (\bibinfo{year}{2010}).

\bibitem[{\citenamefont{Balcan and Vespignani}(2012)}]{JTB201287}
\bibinfo{author}{\bibfnamefont{D.}~\bibnamefont{Balcan}} \bibnamefont{and}
  \bibinfo{author}{\bibfnamefont{A.}~\bibnamefont{Vespignani}},
  \bibinfo{journal}{J. Theor. Biol.} \textbf{\bibinfo{volume}{293}},
  \bibinfo{pages}{87} (\bibinfo{year}{2012}).

\bibitem[{\citenamefont{Barth\'{e}lemy}(2011)}]{PR2011001}
\bibinfo{author}{\bibfnamefont{M.}~\bibnamefont{Barth\'{e}lemy}},
  \bibinfo{journal}{Phys. Rep.} \textbf{\bibinfo{volume}{499}},
  \bibinfo{pages}{1} (\bibinfo{year}{2011}).

\bibitem[{\citenamefont{Nicosia et~al.}(2011)\citenamefont{Nicosia, Bagnoli,
  and Latora}}]{EPL201168009}
\bibinfo{author}{\bibfnamefont{V.}~\bibnamefont{Nicosia}},
  \bibinfo{author}{\bibfnamefont{F.}~\bibnamefont{Bagnoli}}, \bibnamefont{and}
  \bibinfo{author}{\bibfnamefont{V.}~\bibnamefont{Latora}},
  \bibinfo{journal}{Europhys. Lett.} \textbf{\bibinfo{volume}{94}},
  \bibinfo{pages}{68009} (\bibinfo{year}{2011}).

\bibitem[{\citenamefont{Moilanen et~al.}(2009)\citenamefont{Moilanen, Wilson,
  and Possingham}}]{Moi2009}
\bibinfo{author}{\bibfnamefont{A.}~\bibnamefont{Moilanen}},
  \bibinfo{author}{\bibfnamefont{K.}~\bibnamefont{Wilson}}, \bibnamefont{and}
  \bibinfo{author}{\bibfnamefont{H.}~\bibnamefont{Possingham}},
  \emph{\bibinfo{title}{Spatial conservation prioritization}}
  (\bibinfo{publisher}{Oxford University Press}, \bibinfo{address}{London},
  \bibinfo{year}{2009}).

\bibitem[{\citenamefont{Balcan and Vespignani}(2011)}]{NAP2011581}
\bibinfo{author}{\bibfnamefont{D.}~\bibnamefont{Balcan}} \bibnamefont{and}
  \bibinfo{author}{\bibfnamefont{A.}~\bibnamefont{Vespignani}},
  \bibinfo{journal}{Nature Phys.} \textbf{\bibinfo{volume}{7}},
  \bibinfo{pages}{581} (\bibinfo{year}{2011}).

\bibitem[{\citenamefont{Vespignani}(2012)}]{NAP201232}
\bibinfo{author}{\bibfnamefont{A.}~\bibnamefont{Vespignani}},
  \bibinfo{journal}{Nature Phys.} \textbf{\bibinfo{volume}{8}},
  \bibinfo{pages}{32} (\bibinfo{year}{2012}).

\bibitem[{\citenamefont{Nakao and Mikhailov}(2010)}]{NAP2010544}
\bibinfo{author}{\bibfnamefont{H.}~\bibnamefont{Nakao}} \bibnamefont{and}
  \bibinfo{author}{\bibfnamefont{A.~S.} \bibnamefont{Mikhailov}},
  \bibinfo{journal}{Nature Phys.} \textbf{\bibinfo{volume}{6}},
  \bibinfo{pages}{544} (\bibinfo{year}{2010}).

\bibitem[{\citenamefont{Daley and Gani}(1999)}]{Dal1999}
\bibinfo{author}{\bibfnamefont{D.~J.} \bibnamefont{Daley}} \bibnamefont{and}
  \bibinfo{author}{\bibfnamefont{J.}~\bibnamefont{Gani}},
  \emph{\bibinfo{title}{Epidemic Modelling}} (\bibinfo{publisher}{Cambridge
  University Press}, \bibinfo{address}{Cambridge}, \bibinfo{year}{1999}).

\bibitem[{\citenamefont{Hethcote}(2000)}]{SIAM2000599}
\bibinfo{author}{\bibfnamefont{H.~W.} \bibnamefont{Hethcote}},
  \bibinfo{journal}{SIAM Rev.} \textbf{\bibinfo{volume}{42}},
  \bibinfo{pages}{599} (\bibinfo{year}{2000}).

\bibitem[{\citenamefont{Murray}(2002)}]{Mur2002}
\bibinfo{author}{\bibfnamefont{J.~D.} \bibnamefont{Murray}},
  \emph{\bibinfo{title}{Mathematical Biology}}
  (\bibinfo{publisher}{Springer-Verlag}, \bibinfo{address}{Berlin},
  \bibinfo{year}{2002}).

\bibitem[{\citenamefont{Barab\'{a}si and Albert}(1999)}]{SCI99000509}
\bibinfo{author}{\bibfnamefont{A.-L.} \bibnamefont{Barab\'{a}si}}
  \bibnamefont{and} \bibinfo{author}{\bibfnamefont{R.}~\bibnamefont{Albert}},
  \bibinfo{journal}{Science} \textbf{\bibinfo{volume}{286}},
  \bibinfo{pages}{509} (\bibinfo{year}{1999}).

\bibitem[{\citenamefont{Newman}(2002)}]{PRE02016128}
\bibinfo{author}{\bibfnamefont{M.~E.~J.} \bibnamefont{Newman}},
  \bibinfo{journal}{Phys. Rev. E} \textbf{\bibinfo{volume}{66}},
  \bibinfo{pages}{016128} (\bibinfo{year}{2002}).

\bibitem[{\citenamefont{Pastor-Satorras
  et~al.}(2001)\citenamefont{Pastor-Satorras, V\'azquez, and
  Vespignani}}]{PRL01258701}
\bibinfo{author}{\bibfnamefont{R.}~\bibnamefont{Pastor-Satorras}},
  \bibinfo{author}{\bibfnamefont{A.}~\bibnamefont{V\'azquez}},
  \bibnamefont{and}
  \bibinfo{author}{\bibfnamefont{A.}~\bibnamefont{Vespignani}},
  \bibinfo{journal}{Phys. Rev. Lett.} \textbf{\bibinfo{volume}{87}},
  \bibinfo{pages}{258701} (\bibinfo{year}{2001}).

\bibitem[{\citenamefont{Dorogovtsev and Mendes}(2003)}]{Dor2003}
\bibinfo{author}{\bibfnamefont{S.~N.} \bibnamefont{Dorogovtsev}}
  \bibnamefont{and} \bibinfo{author}{\bibfnamefont{J.~F.~F.}
  \bibnamefont{Mendes}}, \emph{\bibinfo{title}{Evolution of Networks: From
  Biological Nets to the Internet and WWW}} (\bibinfo{publisher}{Oxford Univ.
  Press}, \bibinfo{address}{Oxford}, \bibinfo{year}{2003}).

\end{thebibliography}

\end{document}